\documentclass[aps,prd,superscriptaddress,showpacs,preprint]{revtex4}
\usepackage{graphicx, bm}
\usepackage[usenames]{color}

\begin{document}
%\tightenlines
\draft
\title{Probing the electromagnetic dipole moments of the tau-neutrino in the $U(1)_{B-L}$ model at the ILC and CLIC energies}

\author{ A. Llamas-Bugarin\footnote{maria.llamas@fisica.uaz.edu.mx}}
\affiliation{\small Facultad de F\'{\i}sica, Universidad Aut\'onoma de Zacatecas\\
         Apartado Postal C-580, 98060 Zacatecas, M\'exico.\\}

\author{ A. Guti\'errez-Rodr\'{\i}guez\footnote{alexgu@fisica.uaz.edu.mx}}
\affiliation{\small Facultad de F\'{\i}sica, Universidad Aut\'onoma de Zacatecas\\
         Apartado Postal C-580, 98060 Zacatecas, M\'exico.\\}

\author{ M. A. Hern\'andez-Ru\'{\i}z\footnote{mahernan@uaz.edu.mx}}
\affiliation{\small Unidad Acad\'emica de Ciencias Qu\'{\i}micas, Universidad Aut\'onoma de Zacatecas\\
         Apartado Postal C-585, 98060 Zacatecas, M\'exico.\\}

\date{\today}
%\maketitle

\begin{abstract}
% insert abstract here
In this work we study the sensitivity on the anomalous magnetic and electric dipole moments of the tau-neutrino in the
framework of the $SU(2)_L\times U(1)_Y\times U(1)_{B-L}$ electroweak model at future $e^+e^-$ linear colliders as the ILC
and CLIC. For our study we consider the process $e^{+}e^{-}\rightarrow ( Z, Z', \gamma) \to \nu_\tau \bar \nu_\tau \gamma$.
For center-of-mass energies of $\sqrt{s}=1000-3000\hspace{0.8mm}GeV$ and integrated luminosities of
${\cal L}=500-2000\hspace{0.8mm}fb^{-1}$, we derive $95 \% \hspace{1mm}$ C.L. limits on the dipole moments
$|\mu_{\nu_\tau}(\mu_B)| \leq 6.28 \times 10^{-9}$ and $|d_{\nu_\tau}(e cm)| \leq 1.21 \times 10^{-21}$ improve the existing
limits by two or three orders of magnitude. Our study complements other studies on the dipole moments of the tau-neutrino at
hadron and $e^+e^-$ colliders.

\end{abstract}

\pacs{14.60.St, 13.40.Em, 12.15.Mm
%\vspace*{2cm}\noindent  E-mail: $^{1}$alexgu@fisica.uaz.edu.mx
}

\vspace{5mm}

\maketitle

\section{Introduction}

In the Standard Model (SM) \cite{S.L.Glashow,Weinberg,Salam} minimally extended with Dirac neutrino masses, the neutrino magnetic
moment induced by radiative corrections is unobservably small \cite{Fukugita,Robert,Fukugita1},

\begin{equation}
\mu_{\nu_i} = \frac{3m_e G_F}{4\sqrt{2}\pi^2}m_{\nu_i}\simeq 3.1\times 10^{-19}\left(\frac{m_{\nu_i}}{1 \hspace{0.8mm} eV} \right)\mu_B,
\end{equation}

\noindent where $\mu_B=e/2m_e$ is the Bohr Magneton. Current limits on these magnetic moments are several orders of magnitude larger,
so that a magnetic moment close to these limits would indicate a window for probing effects induced by new physics beyond the SM \cite{Fukugita1}.
Similarly, a neutrino electric dipole moment will also point to new physics and will be of relevance in astrophysics and cosmology, as
well as terrestrial neutrino experiments \cite{Cisneros}. Some bounds on the neutrino magnetic moment are shown in Table I.

\begin{table}[!ht]
\caption{Bounds on the neutrino magnetic moment.}
\begin{center}
\begin{tabular}{|c|c|c|c|}

\hline
Experiment/ Method             &  Limit                                            & C. L.             &  Reference\\
\hline\hline
Laboratory experiment Borexino   &  $\mu_{\nu} \leq 5.4\times 10^{-11}\mu_B$       &  $90\%$           & \cite{Borexino}   \\
\hline

Laboratory experiment TEXONO   &  $\mu_{\nu} < 2.9\times 10^{-11}\mu_B$            &  $90\%$           & \cite{Texono}   \\
\hline
Cooling rates of white dwarfs  &  $\mu_\nu \lesssim 10^{-11}\mu_B$                 &  $90\%$           &  \cite{Blinnikov}  \\
\hline
Cooling rates of red giants    &  $\mu_\nu \lesssim 3\times 10^{-12}\mu_B$         &  $90\%$           & \cite{Raffelt}  \\
\hline
Supernova energy loss          & $\mu_\nu \lesssim (1.1-2.7)\times 10^{-12}\mu_B$  &  $90\%$           & \cite{Kuznetsov}  \\
\hline
Absence of high-energy events in the   &  $\mu_\nu \lesssim 10^{-12}\mu_B$         &  $90\%$           &  \cite{Barbieri}  \\
SN1987A neutrino signal                &                                           &                   &                    \\
\hline
Standard model (Dirac mass)    & $\mu_\nu \simeq 3.1\times 10^{-19}(m_\nu/1 \hspace{1mm} eV) \mu_B$ &  & \cite{Fukugita,Robert,Fukugita1}  \\
\hline\hline
\end{tabular}
\end{center}
\end{table}

In the case of the anomalous magnetic moment of the tau-neutrino, the current best limit on $\mu_{\nu_\tau}$ has
been obtained  in the Borexino experiment which explores solar neutrinos. Searches for the magnetic moment of the tau-neutrino
have also been performed in accelerator experiments. The experiment E872 (DONUT) is based at $\nu_\tau e^-, \bar\nu_\tau e^-$
elastic scattering. In the CERN experiment WA-066, a limit on $\mu_{\nu_\tau}$ is obtained on an assumed flux of tau-neutrinos
in the neutrino beam. The L3 collaboration obtain a limit on the magnetic moment of the tau-neutrino from a sample of $e^+e^-$
annihilation events at the $Z$ resonance. Experimental limits on the magnetic moment of the tau-neutrino
are shown in Table II. Others limits on the magnetic moment of the $\mu_{\nu_\tau}$ are reported in the literature
\cite{Gutierrez10,Gutierrez9,Gutierrez12,Gutierrez8,Ruiz1,Data2014,Gutierrez7,Gutierrez11,Gutierrez6,Aydin,Gutierrez5,Gutierrez4,Gutierrez3,
Keiichi,Aytekin,Gutierrez2,Gutierrez1,DELPHI,Escribano,Gould,Grotch,Sahin,Koksal}.

\begin{table}[!ht]
\caption{Experimental limits on the magnetic moment of the tau-neutrino.}
\begin{center}
\begin{tabular}{|c| c| c| c| c|}
\hline
Experiment     &  Method          &  Limit                                      & C. L.  &  Reference\\
\hline
\hline

Borexino       &  Solar neutrino  &  $\mu_{\nu_\tau} < 1.9\times 10^{-10}\mu_B$ & $90 \%$  & \cite{Borexino} \\
\hline
E872 (DONUT)   &  Accelerator $\nu_\tau e^-, \bar\nu_\tau e^-$ & $\mu_{\nu_\tau} < 3.9\times 10^{-7}\mu_B$  & $90 \%$  & \cite{DONUT} \\
\hline
CERN-WA-066    &  Accelerator     & $\mu_{\nu_\tau} < 5.4\times 10^{-7}\mu_B$   & $90 \%$ & \cite{A.M.Cooper} \\
\hline
L3             & Accelerator      & $\mu_{\nu_\tau} < 3.3\times 10^{-6}\mu_B$   & $90 \%$  & \cite{L3} \\
\hline\hline
\end{tabular}
\end{center}
\end{table}

The discovery of CP violation in the decays of neutral kaons \cite{Christenson}, and later in the decays of neutral B mesons \cite{AbeK}
and $D^0$ \cite{LHCb}, shed light on the nature and origin of the violation of this symmetry. The CP violation is one of
the open problems of the SM. For this reason, the measurement of large amounts of CP violation can be
indicative of signs of new physics. The signs of new physics can be analyzed by investigating the electromagnetic dipole moments of the
tau-neutrino such as its magnetic moment (MM) and electric dipole moment (EDM) defined as a source of CP violation.

In the case of the electric dipole moment of the tau-neutrino some theoretical limits are presented in Table III.
Others limits on the $d_{\nu_\tau}$ are reported in the literature
\cite{Gutierrez10,Gutierrez9,Gutierrez12,Gutierrez8,Ruiz1,Gutierrez11,Data2014,Gutierrez7,Gutierrez6,Gutierrez5,Gutierrez4}.

\begin{table}[!ht]
\caption{Theoretical limits on the electric dipole moment of the electron-neutrino, muon-neutrino and the tau-neutrino.}
\begin{center}
\begin{tabular}{|c| c| c| c| c|}
\hline
Particle           &  Model           &  Limit                                                      &    C. L.  &  Reference\\
\hline
\hline

$\nu_{e, \mu}$     &  Model-independent             &  $d_{\nu_e, \nu_\mu} < 2\times 10^{-21}\hspace{0.8mm}e cm$   & $95 \%$  & \cite{Aguila} \\
\hline
$\nu_\tau$         &  Effective Lagrangian approach               &  $d_{\nu_\tau} < 5.2\times 10^{-17}\hspace{0.8mm}e cm$           & $95 \%$  & \cite{Escribano} \\
\hline
$\nu_\tau$         &  Model-independent  &  $d_{\nu_\tau} < O(2 \times 10^{-17} \hspace{0.8mm}e cm)$             & $95 \%$  & \cite{Keiichi} \\
\hline
$\nu_\tau$         & Vector like Multiplets  &  $d_{\nu_\tau} < O( 10^{-18}-10^{-20} \hspace{0.8mm}e cm)$     & $95 \%$  & \cite{Tarek} \\
\hline\hline
\end{tabular}
\end{center}
\end{table}

\newpage

The $U(1)_{B-L}$ model \cite{Buchmuller,Marshak,Mohapatra,Khalil,Khalil1} is one of the simplest extensions of the SM with an
extra $U(1)$ local gauge symmetry \cite{Carlson}, where B and L represent the baryon number and lepton number, respectively.
This B-L symmetry plays an important role in various physics scenarios beyond the SM.
The features that distinguish the $U(1)_{B-L}$ models from other models are the following:
{\bf a)} The gauge $U(1)_{B-L}$ symmetry group is contained in the Grand Unification Theory (GUT) described by a $SO(10)$
group \cite{Buchmuller}. {\bf b)} The scale of the B-L  symmetry breaking is related to the mass scale of the heavy right-handed
Majorana neutrino mass terms and provide the well-known see-saw mechanism \cite{Mohapatra1,Minkowski,Freedman,Yanagida,Ramond}
to explain light left-handed neutrino mass. {\bf c)} The B-L symmetry and the scale of its breaking are tightly connected to the
baryogenesis mechanism through leptogenesis \cite{Fukugita1}. {\bf d)} Another distinctive feature of the $U(1)_{B-L}$ models is the
possibility of the $Z'$ heavy boson decaying into pairs of heavy neutrinos $\Gamma(Z'\to \nu_h\bar\nu_h)$. The model contains an extra
gauge boson $Z'$ corresponding to B-L gauge symmetry and an extra SM singlet scalar (heavy Higgs boson H). These new particles can change
the SM phenomenology significantly and lead to interesting signatures at the current and future colliders such as the Large
Hadron Collider (LHC) \cite{Aad,Chatrchyan}, International Linear Collider (ILC) \cite{Abe,Aarons,Brau,Baer,Asner,Zerwas} and
the Compact Linear Collider (CLIC) \cite{Accomando,Dannheim,Abramowicz}.

The B-L model \cite{Basso,Basso0} is attractive due to its relatively simple theoretical structure. The crucial test of the model
is the detection of the new heavy neutral $(Z')$ gauge boson and the new Higgs boson $(H)$. On the other hand, searches for both
the heavy gauge boson $(Z')$ and the additional heavy neutral Higgs boson $(H)$ predicted by the B-L model are presently being
conducted at the LHC. In this regard, the additional boson $Z'$ of the B-L model has a mass which is given by the relation $M_{Z'}=2v'g'_1$ \cite{Khalil,Khalil1,Basso,Basso0}. This boson $Z'$ interacts with the leptons, quarks, heavy neutrinos and light neutrinos with
interaction strengths proportional to the B-L gauge coupling $g'_1$. The sensitivity limits on the mass $M_{Z'}$  of the boson
$Z'$ of the $U(1)_{B-L}$ model derived for the ATLAS and CMS collaborations are of the order of ${\cal O}(1.83-2.65)\hspace{0.8mm}TeV$ \cite{ATLAS,ATLAS0,CMS0,CMS1,CMS2,CMS3,ATLAS1,ATLAS2,CMS4}. It is noteworthy that future LHC runs at 13-14 $TeV$ could increase the $Z'$
mass bounds to higher values, or evidence may be found of its existence. Precision studies of the $Z'$ properties will require a new linear
collider \cite{Allanach}, which will allow us to perform precision studies of the Higgs sector. We refer the readers to Refs. \cite{Khalil,Khalil1,Basso,Basso0,Basso1,Basso2,Basso3,Basso5,Basso6,Satoshi} for a detailed description of the B-L model.

Our aim in the present paper is to analyze the reaction $e^{+}e^{-}\rightarrow \nu_\tau \bar\nu_\tau \gamma$ in the framework
of the  $U(1)_{B-L}$ model and we attribute an anomalous magnetic moment and an electric dipole moment to a massive tau-neutrino.
It is worth mentioning that at higher $s$, the dominant contribution involves the exchange of the $Z, Z'$ bosons.
The dependence on the magnetic moment $(\mu_{\nu_\tau})$ and the electric dipole moment $(d_{\nu_\tau})$ comes from
the radiation of the photon observed by the neutrino or antineutrino in the final state. However, in order to improve
the limits on the magnetic moment and the electric dipole moment of the tau-neutrino, in our calculation of the process
$e^+e^- \to \nu_\tau \bar\nu_\tau \gamma$ we consider the contribution that involves the exchange of a virtual photon. In
this case, the dependence on the dipole moments comes from a direct coupling to the virtual photon, and the observed photon
is a result of initial-state Bremsstrahlung. The Feynman diagrams which give the most  important contribution to the cross
section are shown in Fig. 1. This process sets limits on the tau-neutrino MM and EDM. In this paper, we take advantage of this
fact to set limits on $\mu_{\nu_{\tau}}$ and $d_{\nu_{\tau}}$ for integrated luminosities of $500-2000 \hspace{0.8mm}fb^{-1}$
and center-of-mass energies between $1000-3000\hspace{0.8mm}GeV$, that is to say in the next generation of linear colliders,
namely, the International Linear Collider (ILC) \cite{Abe} and the Compact Linear Collider (CLIC) \cite{Accomando}.

The L3 Collaboration \cite{L3} evaluated the selection efficiency using detector-simulated $e^{+}e^{-}\rightarrow \nu \bar \nu \gamma (\gamma)$
events, random trigger events, and large-angle $e^{+}e^{-}\rightarrow e^+e^-$ events. From Fig. 1 of Ref. \cite{L3} the process
$e^+e^- \to \nu \bar\nu \gamma$  with $\gamma$ emitted in the initial state is the sole background in the $[44.5^0,135.5^0]$ angular range
(white histogram). From the same figure in this angular interval that is $-0.7<\cos\theta_\gamma <0.7$ we see that only 6 events were found,
this is the real background, not 14 events. In this case a simple method \cite{Data2014,Rick,Bayatian} is that at 1$\sigma$ level
($68\hspace{0.5mm}\% \hspace{2mm}C.L$) for a null signal the number of observed events should not exceed the fluctuation of the estimated
background events: $N = N_B+\sqrt{N_B}$. Of course, this method is good only when $N_B$ is sufficiently large (i.e. when the Poisson distribution
can be  approximated with a gaussian \cite{Data2014,Rick,Bayatian}) but for $N_B > 10$ it is a good approximation. This  means that at $1\sigma, 2\sigma, 3\sigma$ level ($68\hspace{1mm}\%, 90\hspace{1mm}\%, 95\hspace{1mm}\%\hspace{2mm}C.L.$) the limits on the non-standard parameters are found
replacing the equation for the total number of events expected $N=N_B+\sqrt{N_B}$ in the expression $N=\sigma(\mu_{\nu_\tau}, d_{\nu_\tau}){\cal L}$.
The distributions of the photon  energy and the cosine of its polar angle are consistent with SM predictions.

This paper is organized as follows: In Section II, we present the B-L theoretical model. In Sec. III we present the calculation of the process $e^{+}e^{-}\rightarrow \nu_\tau \bar\nu_\tau \gamma$ in the context of the B-L model. Finally, we present our results and conclusions in Sect. IV.

\section{Brief Review of the B-L Theoretical Model}

The solid evidence for the non-vanishing neutrino masses has been confirmed by various neutrino oscillation phenomena and
indicates the evidence of new physics beyond the SM. In the SM, neutrinos are massless due to the absence of right-handed
neutrinos and the exact B-L conservation. The most attractive idea to naturally explain the tiny neutrino masses is the seesaw
mechanism \cite{Minkowski,Freedman,Yanagida,Gellman}, in which the right-handed (RH) neutrinos singlet under the SM gauge
group is introduced. The gauged $U(1)_{B-L}$ model based on the gauge group $SU(3)_C \times SU(2)_L \times U(1)_Y \times U(1)_{B-L}$
\cite{Mohapatra1,Marshak1} is an elegant and simple extension of the SM in which the RH heavy neutrinos are essential both
for anomaly cancelation and preserving gauge invariance. In addition, the mass of RH neutrinos arises associated with the $U(1)_{B-L}$
gauge symmetry breaking. Therefore, the fact that neutrinos are massive indicates that the SM requires extension.

We consider a $SU(3)_C\times SU(2)_L\times U(1)_Y\times U(1)_{B-L}$ model, which is one of the
simplest extensions of the SM \cite{Mohapatra1,Marshak1,Khalil,Khalil1,Basso,Basso1,Basso2,Basso3,Basso5,Basso6,Satoshi},
where $U(1)_{B-L}$, represents the additional gauge symmetry. The gauge invariant Lagrangian of this model is given by

\begin{equation}
{\cal L}={\cal L}_s+{\cal L}_{YM}+{\cal L}_f+{\cal L}_Y,
\end{equation}

\noindent where ${\cal L}_s, {\cal L}_{YM}, {\cal L}_f$ and ${\cal L}_Y$ are the scalar, Yang-Mills, fermion and
Yukawa sector, respectively.

The model consists of one doublet $\Phi$ and one singlet $\chi$ and we briefly describe the lagrangian including
the scalar, fermion and gauge sector, respectively. The Lagrangian for the gauge sector is given by \cite{Ferroglia,Rizzo,Khalil,Basso6},

\begin{equation}
{\cal L}_g=-\frac{1}{4}B_{\mu\nu}B^{\mu\nu}-\frac{1}{4}W^a_{\mu\nu}W^{a\mu\nu}-\frac{1}{4}Z'_{\mu\nu}Z^{'\mu\nu},
\end{equation}

\noindent where $W^a_{\mu\nu}$, $B_{\mu\nu}$ and $Z'_{\mu\nu}$ are the field strength tensors for $SU(2)_L$, $U(1)_Y$
and $U(1)_{B-L}$, respectively.

The Lagrangian for the scalar sector of the model is

\begin{equation}
{\cal L}_s=(D^\mu\Phi)^\dagger(D_\mu\Phi) + (D^\mu\chi)^\dagger(D_\mu\chi)-V(\Phi, \chi),
\end{equation}

\noindent where the potential term is \cite{Basso3},

\begin{equation}
V(\Phi, \chi)=m^2(\Phi^\dagger \Phi)+\mu^2|\chi|^2+\lambda_1(\Phi^\dagger \Phi)^2 + \lambda_2|\chi|^4 + \lambda_3(\Phi^\dagger \Phi)|\chi|^2,
\end{equation}

\noindent with $\Phi$ and $\chi$ as the complex scalar Higgs doublet and singlet fields, respectively.
The covariant derivative is given by \cite{Basso1,Basso2,Basso3}

\begin{equation}
D_\mu=\partial_\mu + ig_st^\alpha G^{\alpha}_\mu + i[gT^aW^a_\mu + g_1YB_\mu + (\tilde gY + g'_1Y_{B-L})B'_\mu ],
\end{equation}

\noindent where $g_s$, $g$, $g_1$ and $g'_1$ are the $SU(3)_C$, $SU(2)_L$, $U(1)_Y$ and $U(1)_{B-L}$ couplings with $t^\alpha$, $T^a$, $Y$
and $Y_{B-L}$ being their respective group generators. The mixing between the two Abelian groups is described by the new coupling $\tilde g$.
The electromagnetic charges on the fields are the same as those of the SM and the $Y_{B-L}$ charges for quarks, leptons and the scalar fields
are given by: $Y^{\mbox{\small quarks}}_{B-L}=1/3$, $Y^{\mbox{\small leptons}}_{B-L}=-1$ with no distinction between generations for ensuring
universality, $Y_{B-L}(\Phi)=0$ and $Y_{B-L}(\chi)=2$ \cite{Khalil,Khalil1,Basso1,Basso2,Basso3} to preserve the gauge invariance of the model,
respectively.

An effective coupling and effective charge such as $g'$ and $Y'$ are usually introduced as $g' Y'= \tilde g Y + g'_1 Y_{B-L}$
and some specific benchmark models \cite{Appelquist,Carena} can be recovered by particular choices of both $\tilde g$ and $g'_1$
gauge couplings at a given scale, generally the electroweak scale. For instance, the pure B-L model is obtain by the condition $\tilde g =0$
$(Y'=Y_{B-L})$ which implies the absence of mixing at the electroweak scale. Other benchmark models of the general parameterisation
are the Sequential Standar Model (SSM), the $U(1)_R$ model and the $U(1)_\chi$ model. The SSM is reproduced by the condition $g'_1=0$
$(Y'=Y)$, and the $U(1)_R$ extension is realised by the condition $\tilde g =-2g'_1$, while the $SO(10)$-inspired $U(1)_\chi$ model
is described by $\tilde g =-\frac{4}{5}g'_1$.

The doublet and singlet scalars are

\begin{eqnarray}
\Phi= \left(
       \begin{array}{c}
        G^{\pm}\\
      \frac{v+\phi^0 +i G_Z}{\sqrt{2}}
      \end{array}
      \right), \hspace{1cm}
\chi= \left(\frac{v'+\phi^{'0} +i  z'}{\sqrt{2}}\right),
\end{eqnarray}

\noindent with $G^{\pm}$, $G_Z$ and $z'$ the Goldstone bosons of $W^{\pm}$, $Z$ and $Z'$, respectively,
while $v\approx 246\hspace{1mm}GeV$ is the electroweak symmetry breaking scale and $v'$ is the B-L symmetry breaking
scale constrained by the electroweak precision measurement data whose value is assumed to be of the order $TeV$.

After spontaneous symmetry breaking, the two scalar fields can be written as,

\begin{eqnarray}
\Phi= \left(
       \begin{array}{c}
        0\\
      \frac{v+\phi^0}{\sqrt{2}}
      \end{array}
      \right), \hspace{1cm}
\chi= \frac{v'+\phi^{'0}}{\sqrt{2}},
\end{eqnarray}

\noindent with $v$ and $v'$ real and positive.

In Table IV, the interactions of $h$ and $H$ with the gauge bosons and scalar are expressed in terms of the parameters of the B-L model.

To determine the mass spectrum of the gauge bosons, we have to expand the scalar kinetic terms as with the SM. We expect that
there exists a massless gauge boson, the photon, while the other gauge bosons become massive. The extension we
are studying is  in the Abelian sector of the SM gauge group, so that the charged gauge bosons $W^\pm$ will have masses given by
their SM expressions related to the $SU(2)_L$ factor only. The other gauge boson masses are not so simple to identify because
of mixing. In fact, analogous to the SM, the fields of definite mass are linear combinations of $B^\mu$, $W^\mu_3$ and $B'^\mu$,
the relation between the neutral gauge bosons ($B^\mu$, $W^\mu_3$ and $B'^\mu$) and the corresponding mass eigenstates
is given by \cite{Basso,Basso0,Basso1,Basso2}

\begin{eqnarray}
\left(
       \begin{array}{c}
        B^{\mu}\\
        W^{3\mu}\\
        B^{'\mu}
      \end{array}
      \right)=\left(
              \begin{array}{c c c}
              \cos\theta_W &   -\sin\theta_W \cos\theta_{B-L} &  \sin\theta_W \sin\theta_{B-L}\\
              \sin\theta_W &  \cos\theta_W \cos\theta_{B-L}   &  -\cos\theta_W \sin\theta_{B-L}\\
                   0       &       \sin\theta_{B-L}           &       \cos\theta_{B-L}
              \end{array}
      \right)
             \left(
       \begin{array}{c}
        A^{\mu}\\
        Z^{\mu}\\
        Z^{'\mu}
      \end{array}
      \right),
\end{eqnarray}

\noindent with $-\frac{\pi}{4} \leq \theta_{B-L} \leq \frac{\pi}{4} $, such that

\begin{equation}
\tan2\theta_{B-L}=\frac{2\tilde g\sqrt{g^2+g^2_1}}{\tilde g^2 + 16(\frac{v'}{v})^2g^{'2}_1-g^2-g^2_1},
\end{equation}

\noindent and the mass spectrum of the gauge bosons is given by

\begin{eqnarray}
M_\gamma&=&0, \nonumber\\
M_{W^\pm}&=&\frac{1}{2}vg, \nonumber\\
M_Z&=& \frac{v}{2}\sqrt{g^2+g^2_1}\sqrt{\frac{1}{2}\biggl(\frac{\tilde g^2+16(\frac{v'}{v})^2 g^{'2}_1 }{g^2+g^2_1}+1\biggr )-\frac{\tilde g}{\sin2\theta_{B-L}\sqrt{g^2 +g^2_1  }}},\\
M_{Z'}&=& \frac{v}{2}\sqrt{g^2+g^2_1}\sqrt{\frac{1}{2}\biggl(\frac{\tilde g^2+16(\frac{v'}{v})^2 g^{'2}_1 }{g^2+g^2_1}+1\biggr )+\frac{\tilde g}{\sin2\theta_{B-L}\sqrt{g^2 +g^2_1  }}}, \nonumber
\end{eqnarray}

\noindent where $M_Z$ and $M_{W^\pm}$ are the SM gauge bosons masses and $M_{Z'}$ is the mass of new neutral gauge boson $Z'$,
which strongly depends on $v'$ and $g'_1$. For $\tilde g =0$, there is no mixing between the new and SM gauge bosons $Z'$ and
$Z$. In this case, the $U(1)_{B-L}$ model is called the pure or minimal model $U(1)_{B-L}$. In this article we consider the case
$\tilde g \neq 0$, which is mostly determined by the other gauge couplings $g_1$ and $g'_1$ \cite{Basso7,Bandyopadhyay,Mansour}. The
electroweak precision measurement data can give stringent constraints on the $Z-Z'$ mixing angle $\theta_{B-L}$ expressed in
Eq. (10) \cite{Schael}.

In the Lagrangian of the $SU(3)_C\times SU(2)_L\times U(1)_Y\times U(1)_{B-L}$ model, the terms for the interactions between
neutral gauge bosons $Z, Z'$ and a pair of fermions of the SM can be written in the form \cite{Khalil,Khalil1,Gutierrez,Gutierrez1,Shi,Francisco}

\begin{equation}
{\cal L}_{NC}=\frac{-ig}{\cos\theta_W}\sum_f\bar f\gamma^\mu\frac{1}{2}(g^f_V- g^f_A\gamma^5)f Z_\mu + \frac{-ig}{\cos\theta_W}\sum_f\bar f\gamma^\mu\frac{1}{2}(g^{'f}_V- g^{'f}_A\gamma^5)f Z'_\mu.
\end{equation}

From this Lagrangian we determine the expressions for the new couplings of the $Z, Z'$ bosons with the SM fermions,
which are given in Table IV. The couplings $g^f_V\hspace{0.8mm}(g^{'f}_V)$ and $g^f_A\hspace{0.8mm}(g^{'f}_A)$ depend on the
$Z-Z'$ mixing angle $\theta_{B-L}$ and the coupling constant $g'_1$ of the B-L interaction. In these couplings, the current bound
on the mixing angle is $|\theta_{B-L}|\leq 10^{-3}$ \cite{Data2014}. In the decoupling limit, when $\theta_{B-L}=0$ and
$g'_1=0$, the couplings of the SM are recovered.

\begin{table}[!ht]
\caption{The new couplings of the $Z, Z'$ bosons with the SM fermions and vector boson, scalar coupling in the B-L model.
$g=e/\sin\theta_W$ and $\theta_{B-L}$ is the $Z-Z'$ mixing angle.}
\begin{center}
 \begin{tabular}{|c|c|}
\hline\hline
Particle       &                 Couplings                       \\
\hline\hline
$f\bar f Z$    &       $g^f_V=T^f_3\cos\theta_{B-L}-2Q_f\sin^2\theta_W\cos\theta_{B-L}+\frac{2g'_1}{g}\cos\theta_W \sin\theta_{B-L},$   \\

               &       $g^f_A=T^f_3\cos\theta_{B-L}$                                                                                 \\
\hline

$f\bar f Z'$   &       $g^{'f}_V=-T^f_3\sin\theta_{B-L}-2Q_f \sin^2\theta_W \sin\theta_{B-L}+\frac{2g'_1}{g}\cos\theta_W \cos\theta_{B-L},$ \\

               &       $g^{'f}_A=-T^f_3\sin\theta_{B-L}$\\

\hline
$Z_\mu Z'_\nu h$    &     $g_{ZZ'h}=2i[\frac{1}{4} v\cos\alpha f(\theta_{B-L},g'_1)- v'\sin\alpha g(\theta_{B-L},g'_1)]g_{\mu\nu}, $   \\
           &    $f(\theta_{B-L},g'_1)=-\sin(2\theta')(g^2_1+g^2_2+g^{'2}_1)-2\cos(2\theta')g´_1\sqrt{g^2_1+g^2_2}$,          \\
           &    $g(\theta_{B-L},g'_1)=\frac{1}{4}\sin(2\theta') g^{'2}_1$                                                    \\
\hline
$Z_\mu Z'_\nu H$    &     $g_{ZZ'H}=2i[\frac{1}{4} v\sin\alpha f(\theta_{B-L},g'_1)+ v'\cos\alpha g(\theta_{B-L},g'_1)]g_{\mu\nu}, $       \\
\hline
$W^-_\mu(p_1)W^+_\nu(p_2)Z'_\rho(p_3)$    &     $g_{W^-W^+ Z'}=-ig\cos\theta_W\sin\theta_{B-L}[(p_1-p_2)_\rho g_{\mu\nu}+(p_2-p_3)_\mu g_{\nu\rho}+(p_3-p_1)_\nu g_{\rho\nu}], $   \\
\hline\hline
\end{tabular}
\end{center}
\end{table}

\section{The decay widths of the $Z'$ boson in the B-L model}

In this section we present the decay widths of the $Z'$ boson \cite{Leike,Langacker,Pavel,Robinet,Barger1,Gutierrez,Francisco}
in the context of the B-L model needed in the calculation of the cross section for the
process $e^+e^- \to \nu_\tau \bar\nu_\tau \gamma$. The decay width of the $Z'$ boson to fermions is given by

\begin{equation}
\Gamma(Z' \to f\bar f)=\frac{2G_F}{3\pi \sqrt{2}}N_c M^2_ZM_{Z'} \sqrt{1-\frac{4M^2_f}{M^2_{Z'}}}\Biggl[(g'^f_V)^2     \biggl\{1+2\biggl(\frac{M^2_f}{M^2_{Z'}}\biggr)\biggr\}+(g'^f_A)^2 \biggl\{1-4\biggl(\frac{M^2_f}{M^2_{Z'}}\biggr)\biggr\}\Biggr],
\end{equation}

\noindent where $N_c$ is the color factor ($N_c=1$ for leptons, $N_c=3$ for quarks) and the couplings
$g'^f_V$ and $g'^f_A$ of the $Z'$ boson with the SM fermions are given in Table IV.

The decay width of the $Z'$ boson to heavy neutrinos is

\begin{equation}
\Gamma(Z' \to \nu_h\bar\nu_h)=\frac{g^{'2}_1}{24\pi}\sin^2\theta_{B-L}M_{Z'}\sqrt{\biggl( 1-\frac{4M^2_{\nu_h}}{M^2_{Z'}}\biggr )^3},
\end{equation}

\noindent where the width given by Eq. (14) implies that the right-handed neutrino must be lighter than half the
$Z'$ mass, $M_{\nu_h} < \frac{M_{Z'}}{2}$, and the conditions under which this inequality holds is for coupled
heavy neutrinos, i.e. with minor mass less than $\frac{M_{Z'}}{2}$. The possibility of the $Z'$ heavy boson decaying
into pairs of heavy neutrinos is certainty one of the most interesting of its features.

The $Z'$ partial decay widths involving vector bosons and the scalar bosons are

\begin{equation}
\Gamma(Z' \to W^+W^-)=\frac{G_F M^2_W}{24\pi\sqrt{2}} \cos^2\theta_W\sin^2\theta_{B-L}M_{Z'}\biggl(\frac{M_{Z'}}{M_Z}\biggr)^4
\sqrt{\biggl(1-4\frac{M^2_W}{M^2_{Z'}}\biggr)^3}
\biggl[1+20\frac{M^2_W}{M^2_{Z'}}+12\frac{M^4_W}{M^4_{Z'}}\biggr],
\end{equation}

\begin{equation}
\Gamma(Z' \to Zh)=\frac{G_F M^2_ZM_{Z'}}{24\pi\sqrt{2}}\sqrt{\lambda_h} \biggl[\lambda_h+12\frac{M^2_Z}{M^2_{Z'}}\biggr]
\biggl[f(\theta_{B-L}, g'_1)\cos\alpha + g(\theta_{B-L}, g'_1)\sin\alpha \biggr]^2,
\end{equation}

\begin{equation}
\Gamma(Z' \to ZH)=\frac{G_F M^2_ZM_{Z'}}{24\pi\sqrt{2}}\sqrt{\lambda_H} \biggl[\lambda_H+12\frac{M^2_Z}{M^2_{Z'}}\biggr]
\biggl[f(\theta_{B-L}, g'_1)\sin\alpha - g(\theta_{B-L}, g'_1)\cos\alpha \biggr]^2,
\end{equation}

\noindent where

\begin{eqnarray}
\lambda_{h, H}\biggl(1, \frac{M^2_Z}{M^2_{Z'}}, \frac{M^2_{h, H}}{M^2_{Z'}}\biggr)&=&1+\biggl(\frac{M^2_Z}{M^2_{Z'}}\biggr)^2+\biggl(\frac{M^2_{h, H}}{M^2_{Z'}}\biggr)^2-2\biggl(\frac{M^2_Z}{M^2_{Z'}}\biggr)
-2\biggl(\frac{M^2_{h, H}}{M^2_{Z'}}\biggr)-2\biggl(\frac{M^2_Z}{M^2_{Z'}}\biggr)\biggl(\frac{M^2_{h, H}}{M^2_{Z'}}\biggr),\nonumber\\
f(\theta_{B-L}, g'_1)&=&\biggl(1+\frac{v^2g'^2_1}{4M^2_Z}\biggr)\sin(2\theta_{B-L})+\biggl(\frac{vg'_1}{M_Z}\biggr)\cos(2\theta_{B-L}),\\
g(\theta_{B-L}, g'_1)&=&\biggl(\frac{vv'}{4M^2_Z}\biggr)g'^2_1\sin(2\theta_{B-L}).\nonumber
\end{eqnarray}

\noindent

In the B-L model, the heavy gauge boson mass $M_{Z'}$ satisfies the relation $M_{Z'}=2v'g'_1$ \cite{Khalil,Khalil1,Basso,Basso0,Basso1,Basso2},
and considering the most recent limit from $\frac{M_{Z'}}{g'_1}\geq 6.9\hspace{0.8mm}TeV$ \cite{Heek,Cacciapaglia,Carena},
it is possible to obtain a direct bound on the B-L breaking scale $v'$. In our next numerical calculation, we will take
$v'=3.45\hspace{0.8mm}TeV$, while $\alpha=\frac{\pi}{9}$ for the $h-H$ mixing angle in correspondence with Refs. \cite{Aad,Chatrchyan,Basso4,Khalil}.

\vspace{3mm}

\section{The Total Cross Section}

\vspace{3mm}

In this section we calculate the total cross section for the reaction $e^{+}e^{-}\rightarrow \nu_\tau \bar\nu_\tau \gamma$. The respective
transition amplitudes are thus given by

%\newpage

\begin{eqnarray}
{\cal M}_{1}&=&\frac{-g^{2}}{4\cos^{2}\theta_{W}(l^{2}-m^{2}_{\nu})}
\Bigl[\bar u(p_{3})\Gamma^{\alpha}(\l\llap{/}+m_{\nu})\gamma^{\beta}(g^\nu_{\mbox v}- g^\nu_{A}\gamma_{5})v(p_{4})\Bigr]\\
&\times&\frac{(g_{\alpha\beta}-p_{\alpha}p_{\beta}/M^{2}_{Z})}{\Bigl[(p_{1}+p_{2})^{2}-M^{2}_{Z}-i M_Z\Gamma_Z\Bigr]}\Bigl[\bar
u(p_{2})\gamma^{\alpha}(g^e_{\mbox v}-g^e_{A}\gamma_{5})v(p_{1})\Bigr]\epsilon^{\lambda}_{\alpha},\nonumber
\end{eqnarray}

\begin{eqnarray}
{\cal
M}_{2}&=&\frac{-g^{2}}{4\cos^{2}\theta_{W}(l^{'2}-m^{2}_{\nu})}
\Bigl[\bar u(p_{3})\gamma^{\beta} (g^\nu_{\mbox v}- g^\nu_{A}\gamma_{5})(l\llap{/}'+m_{\nu})\Gamma^{\alpha} v(p_{4})\Bigr]\\
&\times&\frac{(g_{\alpha\beta}-p_{\alpha}p_{\beta}/M^{2}_{Z})}{\Bigl[(p_{1}+p_{2})^{2}-M^{2}_{Z}-i M_Z\Gamma_Z\Bigr]}\Bigl[\bar
u(p_{2})\gamma^{\alpha}(g^e_{\mbox v}-g^e_{A}\gamma_{5})v(p_{1})\Bigr]\epsilon^{\lambda}_{\alpha},\nonumber
\end{eqnarray}

\begin{eqnarray}
{\cal M}_{3}&=&\frac{-g^{2}}{4\cos^{2}\theta_{W}(r^{2}-m^{2}_{\nu})}
\Bigl[\bar u(p_{3})\Gamma^{\alpha}(r\llap{/}+m_{\nu})\gamma^{\beta}(g^{'\nu}_{\mbox v}- g^{'\nu}_{A}\gamma_{5})v(p_{4})\Bigr]\\
&\times&\frac{(g_{\alpha\beta}-p_{\alpha}p_{\beta}/M^{2}_{Z'})}{\Bigl[(p_{1}+p_{2})^{2}-M^{2}_{Z'}-i M_{Z'}\Gamma_{Z'}\Bigr]}\Bigl[\bar
u(p_{2})\gamma^{\alpha}(g^{'e}_{\mbox v}-g^{'e}_{A}\gamma_{5})v(p_{1})\Bigr]\epsilon^{\lambda}_{\alpha},\nonumber
\end{eqnarray}

\begin{eqnarray}
{\cal M}_{4}&=&\frac{-g^{2}}{4\cos^{2}\theta_{W}(r^{'2}-m^{2}_{\nu})}
\Bigl[\bar u(p_{3})\gamma^{\beta} (g^{'\nu}_{\mbox v}- g^{'\nu}_{A}\gamma_{5})(r\llap{/}'+m_{\nu})\Gamma^{\alpha} v(p_{4})\Bigr]\\
&\times&\frac{(g_{\alpha\beta}-p_{\alpha}p_{\beta}/M^{2}_{Z'})}{\Bigl[(p_{1}+p_{2})^{2}-M^{2}_{Z'}-i M_{Z'}\Gamma_{Z'}\Bigr]}\Bigl[\bar
u(p_{2})\gamma^{\alpha}(g^{'e}_{\mbox v}-g^{'e}_{A}\gamma_{5})v(p_{1})\Bigr]\epsilon^{\lambda}_{\alpha},\nonumber
\end{eqnarray}

\begin{eqnarray}
{\cal M}_{5}&&=\frac{e^{2}}{(k^{2}-m^{2}_{e})}\Bigl[\bar u(p_{3})\Gamma^{\alpha}v(p_{4})\Bigr]
\frac{g_{\alpha\beta}}{(p_{1}+p_{2})^{2}}\Bigl[\bar u(p_{2})\gamma^{\alpha}(k\llap{/}+m_e)\gamma^{\beta}
v(p_{1})\Bigr]\epsilon^{\lambda}_{\alpha},
\end{eqnarray}

\noindent and

\begin{eqnarray}
{\cal M}_{6}&&=\frac{e^{2}}{(k^{'2}-m^{2}_{e})}\Bigl[\bar u(p_{3})\Gamma^{\alpha}v(p_{4})\Bigr]
\frac{g_{\alpha\beta}}{(p_{1}+p_{2})^{2}}\Bigl[\bar u(p_{2})\gamma^{\beta}
(k\llap{/}'+m_e)\gamma^{\alpha} v(p_{1})\Bigr]\epsilon^{\lambda}_{\alpha},
\end{eqnarray}

\noindent where the most general expression consistent with Lorentz and electromagnetic gauge invariance,
for the tau-neutrino electromagnetic vertex may be parameterized in terms of four form factors:

\begin{equation}
\Gamma^{\alpha}=eF_{1}(q^{2})\gamma^{\alpha}+\frac{ie}{2m_{\nu_\tau}}F_{2}(q^{2})\sigma^{\alpha
\mu}q_{\mu}+eF_3(q^2)\gamma_5\sigma^{\alpha\mu}q_\mu +eF_4(q^2)\gamma_5(\gamma^\mu q^2-q\llap{/}q^\mu),
\end{equation}

\noindent where $e$ is the charge of the electron, $m_{\nu_\tau}$ is the mass of the tau-neutrino, $q^\mu$ is the photon momentum,
and $F_{1, 2, 3, 4}(q^2)$ are the electromagnetic form factors of the neutrino, corresponding to charge radius, MM, EDM and anapole
moment (AM), respectively, at $q^2=0$ \cite{Escribano,Vogel,Bernabeu1,Bernabeu2,Dvornikov,Giunti,Broggini}, while $\epsilon^\lambda_\alpha$
is the polarization vector of the photon. $l, r (k)$ and $l', r'(k')$ stand for the momentum of the virtual neutrino (electron) and
antineutrino (positron) respectively. The form factors corresponding to charge radius and the anapole moment, do not concern us here.

The MM and EDM give a contribution to the total cross section for the process $e^{+}e^{-}\rightarrow \nu_\tau \bar\nu_\tau \gamma$ of the form:

\begin{eqnarray}
\sigma_{Tot}(e^{+}e^{-}\rightarrow \nu_\tau\bar\nu_\tau\gamma)&=&\int \frac{\alpha^2}{96\pi}\left(\kappa^2\mu^2_B+d^2_{\nu_\tau}\right) \nonumber\\
&\times& \biggl\{ 4 \left[\frac{(g^e_{\mbox v})^2 + (g^e_A)^2}{x^2_W(1-x_W)^2}\right]
\left[\frac{ ( (g^{\nu}_{\mbox v})^2 + (g^{\nu}_A)^2 )(s-2\sqrt{s}E_\gamma) +
(g^\nu_A)^2 E^2_\gamma\sin^2\theta_\gamma}{(s-M^2_{Z})^2+M^2_{Z}\Gamma^2_{Z}}\right]\Bigr. \nonumber\\
&+& 4 \left[\frac{(g^{'e}_{\mbox v})^2 + (g^{'e}_A)^2}{x^2_W(1-x_W)^2}\right]
\left[\frac{ ( (g^{'\nu}_{\mbox v})^2 + (g^{'\nu}_A)^2 )(s-2\sqrt{s}E_\gamma) +
(g^{'\nu}_A)^2 E^2_\gamma\sin^2\theta_\gamma}{(s-M^2_{Z'})^2+M^2_{Z'}\Gamma^2_{Z'}}\right]\Bigr. \nonumber\\
&+&32\left[\frac{s-2\sqrt{s}E_\gamma+2E^2_\gamma- E^2_\gamma\sin^2\theta_\gamma}{sE^2_\gamma\sin^2\theta_\gamma}\right] \nonumber\\
&+&6 \left[\frac{ (g^e_{\mbox v} g^{'e}_{\mbox v} + g^e_A g^{'e}_A)}{x^2_W(1-x_W)^2}\right]
\left[\frac{ (s-M^2_{Z})(s-M^2_{Z'}) + M_{Z}M_{Z'}\Gamma_{Z}\Gamma_{Z'}}
{ [(s-M^2_{Z})^2+M^2_{Z}\Gamma^2_{Z}][(s-M^2_{Z'})^2+M^2_{Z'}\Gamma^2_{Z'}]}\right]\Bigr. \nonumber\\
&\times& \left[ (g^e_{\mbox v} g^{'e}_{\mbox v} + g^e_A g^{'e}_A)(s-2\sqrt{s}E_\gamma) +
(g^\nu_A g^{'\nu}_A)E^2_\gamma\sin^2\theta_\gamma\right]
\Bigr.\biggr\}{E_\gamma dE_\gamma d\cos\theta_\gamma}, \nonumber\\
\end{eqnarray}

\noindent where $x_{W}\equiv \sin^{2}\theta_{W}$ and $E_{\gamma}$, $\cos\theta_{\gamma}$
are the energy and the opening angle of the emitted photon.

The expression given in Eq. (26) corresponds to the total cross section with the exchange of the $Z, Z', \gamma$ bosons.
The SM expression for the cross section of the reaction $e^+e^- \to \nu_\tau \bar\nu_\tau \gamma$ can be obtained
in the decoupling limit when $\theta_{B-L}= 0$, $g'_1=0$ and $\alpha=0$. In this case, the terms that depend on
$\theta_{B-L}$, $g'_1$ and $\alpha$ in Eq. (26) are zero and Eq. (26) is reduced to the expression given in
Ref. \cite{Gould} for the standard model minimally extended to include massive Dirac neutrinos.

\section{Results and Conclusions}

\vspace{3mm}

In order to evaluate the integral of the total cross section as a function of the parameters of the model, that is to say,
$\mu_{\nu_\tau}$ and $d_{\nu_\tau}$ we require cuts on the photon angle and energy to avoid divergences when the integral
is evaluated at the important intervals of each experiment. We integrate over $\theta_\gamma$ from $44.5^o$ to $135.5^o$
and $E_\gamma$ from 15 $GeV$ to 100 $GeV$. Using the following values for numerical computation \cite{Data2014}: $\sin^2\theta_W=0.23126\pm 0.00022$,
$m_\tau=1776.82\pm 0.16\hspace{0.8mm}MeV$, $m_b=4.6\pm 0.18\hspace{0.8mm}GeV$, $m_t=172\pm 0.9\hspace{0.8mm}GeV$,
$M_{W^\pm}=80.389\pm 0.023\hspace{0.8mm}GeV$, $M_Z=91.1876\pm 0.0021\hspace{0.8mm}GeV$, $\Gamma_Z=2.4952\pm 0.0023\hspace{0.8mm}GeV$,
$M_h=125\pm 0.4\hspace{0.8mm}GeV$, $M_H=500\hspace{0.8mm}GeV$ and considering the most recent limit from \cite{Heek,Cacciapaglia,Carena}:

\begin{equation}
\frac{M_{Z'}}{g'_1}\geq 6.9\hspace{0.8mm}TeV,
\end{equation}

\noindent it is possible to obtain a direct bound on the B-L breaking scale $v'$ and take $v'=3.45\hspace{0.8mm}TeV$
and $\alpha=\frac{\pi}{9}$. In our numerical analysis, we obtain the total cross section $\sigma_{Tot}=\sigma_{Tot}(\mu_{\nu_\tau}, d_{\nu_\tau},\sqrt{s},
M_{Z'}, g'_1, \theta_{B-L}, \alpha)$. Thus, in our numerical computation, we will assume $\sqrt{s}$, $M_{Z'}$, $g'_1$, $\theta_{B-L}$
and $\alpha$ as free parameters.

As was discussed in Refs. \cite{Gould,L3,Barnett,Feldman}, $N\approx \sigma_{Tot}(\mu_{\nu_\tau}, d_{\nu_\tau},\sqrt{s},
M_{Z'}, g'_1, \theta_{B-L}, \alpha){\cal L}$, where $N=N_B+\sqrt{N_B}$ is the total number of $e^{+}e^{-}\rightarrow \nu_\tau\bar\nu_\tau\gamma$
events expected at $1\sigma, 2\sigma, 3\sigma$ level as is mentioned in the introduction and ${\cal L}= 500-2000$\hspace{1mm}$fb^{-1}$
according to the data reported by the ILC and CLIC Refs. \cite{Abe, Accomando}. Taking this into consideration, we can obtain a limit
for the tau-neutrino magnetic moment with $d_{\nu_\tau}=0$.

\begin{table}[!ht]
\caption{Bounds on the $\mu_{\nu_\tau}$ magnetic moment and $d_{\nu_\tau}$ electric dipole
moment for $\sqrt{s}=1000, 2000, 3000\hspace{0.8mm}GeV$ and
${\cal L}=500, 1000, 2000\hspace{0.8mm}fb^{-1}$ at $1\sigma$, $2\sigma$ and $3\sigma$.}
\begin{center}
 \begin{tabular}{ccc}
\hline\hline
\multicolumn{3}{c}{${\cal L}=500,\hspace{0.8mm} 1000,\hspace{0.8mm} 2000\hspace{0.8mm}fb^{-1}$}\\
\hline\hline
\multicolumn{3}{c}{\hspace{5mm} ${\sqrt{s}}=1000\hspace{0.8mm}GeV$; \hspace{0.8mm} $M_{Z'}=1000\hspace{0.8mm}GeV, \hspace{0.8mm} g'_1=0.145$ }\\
\hline
\cline{1-3} C. L.          & $|\mu_{\nu_\tau}(\mu_B)|$        & $|d_{\nu_\tau}(e cm)|$ \\
\hline
$1\sigma$                   & ( 3.11, 2.20, 1.55)$\times 10^{-8}$ & \hspace{4mm} ( 6.01, 4.25, 3.00)$\times 10^{-19}$ \\
$2\sigma$                   & ( 3.53, 2.50, 1.76)$\times 10^{-8}$ & \hspace{2mm} ( 6.82, 4.82, 3.41)$\times 10^{-19}$ \\
$3\sigma$                   & ( 3.91, 2.75, 1.95)$\times 10^{-8}$ & \hspace{2mm} ( 7.55, 5.34, 3.77)$\times 10^{-19}$ \\
\hline\hline
\multicolumn{3}{c}{\hspace{5mm} ${\sqrt{s}}=2000\hspace{0.8mm}GeV$; \hspace{0.8mm} $M_{Z'}=2000\hspace{0.8mm}GeV, \hspace{0.8mm} g'_1=0.290$ }\\
 \hline
 \cline{1-3} C. L.          & $|\mu_{\nu_\tau}(\mu_B)|$        & $|d_{\nu_\tau}(e cm)|$ \\
\hline
$1\sigma$                   & ( 1.51, 1.07 )$\times 10^{-8}$,    7.57$\times 10^{-9}$ & \hspace{4mm} ( 2.92, 2.06, 1.46)$\times 10^{-19}$ \\
$2\sigma$                   & ( 1.72,   1.21 )$\times 10^{-8}$,  8.60$\times 10^{-9}$ & \hspace{2mm} ( 3.31, 2.34, 1.65)$\times 10^{-19}$ \\
$3\sigma$                   & ( 1.90, 1.34)$\times 10^{-8}$,     9.52$\times 10^{-9}$ & \hspace{2mm} ( 3.67, 2.59, 1.83 )$\times 10^{-19}$ \\
\hline\hline
\multicolumn{3}{c}{\hspace{5mm} $\sqrt{s}=3000\hspace{0.8mm}GeV$, \hspace{0.8mm} $M_{Z'}=3000\hspace{0.8mm}GeV, \hspace{0.8mm} g'_1=0.435$ }\\
 \hline
 \cline{1-3} C. L.          & $|\mu_{\nu_\tau}(\mu_B)|$        & $|d_{\nu_\tau}(e cm)|$ \\
\hline
$1\sigma$                   & $1.00 \times 10^{-8}$, $( 7.07,  5.00) \times 10^{-9}$ & \hspace{4mm} ( 1.93, 1.36)$\times 10^{-19}$, 9.65$\times 10^{-20}$ \\
$2\sigma$                   & $ 1.13 \times 10^{-8}$, $( 8.03, 5.68) \times 10^{-9}$ & \hspace{2mm} ( 2.19, 1.55, 1.09)$\times 10^{-19}$ \\
$3\sigma$                   & $ 1.25 \times 10^{-8}$, $( 8.89, 6.28 ) \times 10^{-9}$ & \hspace{2mm} ( 2.42, 1.71, 2.21 )$\times 10^{-19}$ \\
\hline\hline
\end{tabular}
\end{center}
\end{table}

As an indicator of the order of magnitude on the dipole moments, we present the bounds obtained on the $\mu_{\nu_\tau}$ magnetic
moment and $d_{\nu_\tau}$ electric dipole moment in Table V for several center-of-mass energies $\sqrt{s}=1000, 2000, 3000\hspace{0.8mm}GeV$,
integrated luminosity ${\cal L}=500, 1000, 2000\hspace{0.8mm}fb^{-1}$ and heavy gauge boson masses $M_{Z'}=1000, 2000, 3000\hspace{0.8mm}GeV$
with $g'_1=0.145, 0.290, 0.435$ at $1\sigma$, $2\sigma$ and $3\sigma$, respectively. It is worth mentioning that the values reported
in Table V for the dipole moments are determined while preserving the relationship between $M_{Z'}$ and $g'_1$ given in Eq. (27). This
relationship will always remain throughout the article. We observed that the results obtained in Table V are better than those reported in the
literature \cite{Gould,Grotch,L3,Escribano,DELPHI,Gutierrez1,Gutierrez2,
Gutierrez3,Gutierrez4,Gutierrez5,Gutierrez6,Gutierrez7,Aydin,Aytekin,Keiichi,A.M.Cooper}.

\begin{table}[!ht]
\caption{Bounds on the $\mu_{\nu_\tau}$ magnetic moment and $d_{\nu_\tau}$ electric dipole
moment for $\sqrt{s}=1000, 2000, 3000\hspace{0.8mm}GeV$ and
${\cal L}=500, 1000, 2000\hspace{0.8mm}fb^{-1}$ at $1\sigma$, $2\sigma$ and $3\sigma$.}
\begin{center}
 \begin{tabular}{ccc}
\hline\hline
\multicolumn{3}{c}{${\cal L}=500,\hspace{0.8mm} 1000,\hspace{0.8mm} 2000\hspace{0.8mm}fb^{-1}$}\\
\hline\hline
\multicolumn{3}{c}{\hspace{5mm} ${\sqrt{s}}=1000\hspace{0.8mm}GeV$ }\\
\hline
\cline{1-3} C. L.          & $|\mu_{\nu_\tau}(\mu_B)|$        & $|d_{\nu_\tau}(e cm)|$ \\
\hline
$1\sigma$                   & ( 2.22, 1.57, 1.11)$\times 10^{-7}$ & \hspace{4mm} ( 4.29, 3.03, 2.14)$\times 10^{-18}$ \\
$2\sigma$                   & ( 2.52, 1.78, 1.26)$\times 10^{-7}$ & \hspace{2mm} ( 4.87, 3.44, 2.43)$\times 10^{-18}$ \\
$3\sigma$                   & ( 2.79, 1.97, 1.39)$\times 10^{-7}$ & \hspace{2mm} ( 5.39, 3.81, 2.69)$\times 10^{-18}$ \\
\hline\hline
\multicolumn{3}{c}{\hspace{5mm} ${\sqrt{s}}=2000\hspace{0.8mm}GeV$ }\\
 \hline
 \cline{1-3} C. L.          & $|\mu_{\nu_\tau}(\mu_B)|$        & $|d_{\nu_\tau}(e cm)|$ \\
\hline
$1\sigma$                   &  1.08 $\times 10^{-7}$,   ( 7.64, 5.40) $\times 10^{-8}$ & \hspace{4mm} ( 2.08, 1.47, 1.04)$\times 10^{-18}$ \\
$2\sigma$                   &  1.22 $\times 10^{-7}$,   ( 8.68, 6.14) $\times 10^{-8}$ & \hspace{2mm} ( 2.36, 1.67, 1.18)$\times 10^{-18}$ \\
$3\sigma$                   &  1.35 $\times 10^{-7}$,   ( 9.61, 6.79) $\times 10^{-8}$ & \hspace{2mm} ( 2.62, 1.85, 1.31 )$\times 10^{-18}$ \\
\hline\hline
\multicolumn{3}{c}{\hspace{5mm} $\sqrt{s}=3000\hspace{0.8mm}GeV$ }\\
 \hline
 \cline{1-3} C. L.          & $|\mu_{\nu_\tau}(\mu_B)|$        & $|d_{\nu_\tau}(e cm)|$ \\
\hline
$1\sigma$                   &  $( 7.14, 5.05, 3.57 ) \times 10^{-8}$ & \hspace{4mm}   1.37$\times 10^{-18}$, ( 9.74,    6.88 $\times 10^{-19}$ \\
$2\sigma$                   &  $( 8.11, 5.73, 4.05 ) \times 10^{-8}$ & \hspace{2mm}   ( 1.56, 1.10)$\times 10^{-18}$,   7.82 $\times 10^{-19}$ \\
$3\sigma$                   &  $( 8.97, 6.34, 4.48 ) \times 10^{-9}$ & \hspace{2mm}   ( 1.73, 1.22 )$\times 10^{-18}$,  8.65 $\times 10^{-19}$ \\
\hline\hline
\end{tabular}
\end{center}
\end{table}

The previous analysis and comments can readily be translated to the EDM of the $\tau$-neutrino with $\mu_{\nu_\tau}=0$.
The resulting limits for the EDM as a function of $\sqrt{s}, M_{Z'}$ and $g'_1$ are shown in Table V.

In the case of the standard model minimally extended \cite{Gould}, i.e. in the decoupling limit when $\theta_{B-L}= 0$, $g'_1=0$
and $\alpha=0$, the bounds generated on the dipole moments are given in Table VI. These bounds are weaker than those obtained with
the $U(1)_{B-L}$ model.

The vector and axial-vector $e^+ e^- Z$ couplings $g^e_V$ and $g^e_A$ which depend on $g'_1$ and $\theta_{B-L}$ are given
in Table IV. To see the dependence of $g^e_V$ and $g^e_A$ on the parameters of the model we plot the relative correction
$\frac{\delta g^e_V}{(g^e_V)_{SM}}=\frac{(g^e_V)_{B-L} - (g^e_V)_{SM}}{ (g^e_V)_{SM} }$ and
$\frac{\delta g^e_A}{(g^e_A)_{SM}}=\frac{(g^e_A)_{B-L} - (g^e_A)_{SM}}{ (g^e_A)_{SM} }$ as a function of $(g'_1, \theta_{B-L})$
in Fig. 2. From the top panel, we can see that the absolute value of the relative correction $\frac{\delta g^e_V}{(g^e_V)_{SM}}$
increases when the parameter $g'_1$ increases and is almost independent of the mixing angle $\theta_{B-L}$. However, the absolute
value of $\frac{\delta g^e_V}{(g^e_V)_{SM}}$ is in the ranges from $10\%-70\%$ in most of the parameter space. In the bottom panel,
we present the relative correction $\frac{\delta g^e_A}{(g^e_A)_{SM}}$ as a function of $g'_1$ and $\theta_{B-L}$. Here it is shown
that the absolute value of $\frac{\delta g^e_A}{(g^e_A)_{SM}}$ increases when the parameter $g'_1$ increases and is almost independent
of the mixing angle $\theta_{B-L}$. For $g'_1=1$, the absolute value of $\frac{\delta g^e_A}{(g^e_A)_{SM}}$ is in the range of $4\%$.
We find that the relative change in $g^e_V$ is much greater than that for $g^e_A$ for the values of the free parameters $g'_1$ and
$\theta_{B-L}$ near the endpoints. We conclude that the deviations of the couplings $g^e_V$ and $g^e_A$ from its SM value are relatively
large in the parameter space $(g'_1, \theta_{B-L})$.

In Fig. 3 we present the total decay width of the $Z'$ boson as a function of $M_{Z'}$ and the new $U(1)_{B-L}$ gauge
coupling $g'_1$, respectively, with the other parameters held fixed to three different values and $\theta_{B-L}=10^{-3}$.
From the top panel, we see that the total width of the $Z'$ new gauge boson varies from very few to hundreds of $GeV$ over
a mass range of $1000\hspace{0.8mm}GeV \leq M_{Z'} \leq 3500\hspace{0.8mm}GeV$, depending on the value of $g'_1$, when
$g'_1=0.145, 0.290, 0.435$, respectively. In the case of the bottom panel, a similar behavior is obtained in the range
$0 \leq g'_1 \leq 1$ and depends on the value $M_{Z'}=1000, 2000, 3000\hspace{0.8mm}GeV$. In both figures a clear dependence
is observed on the parameters of the $U(1)_{B-L}$ model.

Figure 4 shows the total cross section for $e^{+}e^{-}\rightarrow \nu_\tau\bar\nu_\tau\gamma$ as a function of the center-of-
mass energy $\sqrt{s}$ and different values representative of the magnetic moment, which are reported in the literature, that
is to say, $\mu_{\nu_\tau}= 3.3\times 10^{-6}\mu_B\hspace{1mm}\mbox{(L3)}, 5.4\times 10^{-7}\mu_B\hspace{1mm}\mbox{(BEBC (CERN))},
2.75\times 10^{-8}\mu_B\hspace{1mm}\mbox{(Table V)}$ with $M_{Z'}=3000\hspace{0.8mm}GeV$ and $g'_1=0.435$. Starting from a
center-of-mass energy of the order of the $Z$ mass, a minimum around $\sqrt{s}\simeq 100\hspace{1mm}GeV$ occurs due to the SM $Z$-boson
resonance tail on the high energies. For different values of the parameter $\mu_{\nu_\tau}$ the shape of the curves does not change
and there is only a shift of these depending on the value of the magnetic moment.

The dependence of the sensitivity limits of the magnetic moment $\mu_{\nu_\tau}$ with respect to the collider luminosity ${\cal L}$ for
three different values of the center-of-mass energy, $\sqrt{s}= 1000, 2000, 3000\hspace{1mm}GeV$, heavy gauge boson mass of
$M_{Z'}= 1000, 2000, 3000$\hspace{1mm}$GeV$ and $g'_1=0.145, 0.290, 0.435$, respectively, is presented in Fig. 5. The figure clearly shows
a strong dependence of $\mu_{\nu_\tau}$ with respect to ${\cal L}$ and the parameters of the $U(1)_{B-L}$ model. In addition, the spacing
between the curves are broader for larger $g'_1$ values, as the total width of the $Z'$ boson increases with $g'_1$, as shown in figure 3.
Finally, in order to see how the total cross section $e^+e^- \to \nu_\tau \bar\nu_\tau \gamma$ change with respect to the dipole moments
$\mu_{\nu_\tau}$ and $d_{\nu_\tau}$ we give a 3D plot as shown in Fig. 6. In this figure we consider $M_{Z'}= 3000$\hspace{1mm}$GeV$ and
$g'_1= 0.435$ in correspondence with Eq. (27).

It is worth mentioning that by reversing the process, we can obtain specific predictions on the $U(1)_{B-L}$ models from the expression of the
scattering cross section of the process $e^+e^-\to \nu_\tau \bar \nu_\tau\gamma$. Predictions about the models can be obtained by using the upper
bound on the $\nu_\tau$ magnetic moment reported in the literature by the L3 Collaboration as an input, which maximize the total cross section, namely $\mu_{\nu_\tau}=3.3\times 10^{-6}\mu_B$ $(90\% \hspace{1mm}C.L.)$ \cite{L3}, and using the data obtained by the ALEPH Collaboration $\sigma=(3.09\pm 0.234)$ $pb$ Ref. \cite{Heister,Taylor} for the cross section, where the error is statistical.

In conclusion, we have found that the process $e^{+}e^{-}\rightarrow \nu_\tau\bar\nu_\tau\gamma$ in the context of the
standard model minimally extended to include massive Dirac neutrino at the high energies and luminosities expected at
the ILC/CLIC colliders can be used to probe for bounds on the magnetic moment $\mu_{\nu_\tau}$ and electric dipole moment
$d_{\nu_\tau}$. In particular, we can appreciate that the $95 \%\hspace{1mm}$ C.L. sensitivity limits expected for the
magnetic moment at $1000-3000\hspace{0.8mm}GeV$ center-of-mass energies  already can provide proof of these bounds of order
$10^{-8}-10^{-9}$, that is to say, 2-3 orders of magnitude better than those reported in the literature, see Table II and refs. \cite{Gutierrez10,Gutierrez9,Gutierrez12,Gutierrez8,Ruiz1,Gutierrez1,Data2014,Gutierrez7,Gutierrez6,Aydin,Gutierrez5,Gutierrez4,Gutierrez3,Keiichi,Aytekin,Gutierrez2,
Gutierrez1,DELPHI,Escribano,Gould,Grotch}. Our results in Table V compare favorably with the limits
obtained by the L3 Collaboration \cite{L3}, and with other limits reported in the literature
\cite{Gould,Grotch,L3,Escribano,DELPHI,Gutierrez1,Gutierrez2,
Gutierrez3,Gutierrez4,Gutierrez5,Gutierrez6,Gutierrez7,Aydin,Aytekin,Keiichi,A.M.Cooper}.

In the case of the electric dipole moment the $95\%\hspace{1mm}$ C.L. sensitivity limits at
$1000-3000\hspace{0.8mm}GeV$ center-of-mass energies and integrated luminosities of $2000\hspace{1mm}fb^{-1}$
can provide proof of these bounds of order $10^{-19}-10^{-20}$, that is to say, are improved by 2-3 orders
of magnitude than those reported in the literature, see Table III and refs. \cite{Gutierrez10,Gutierrez9,Gutierrez12,Gutierrez8,Ruiz1,Data2014,Gutierrez7,Gutierrez6,Gutierrez5,Gutierrez4}.

The above results do not appear outside the realm of detection in future experiment with improved sensitivity.  In addition,
the analytical and numerical results for the cross section could be of relevance for the scientific community. Further, the
results above could have possible astrophysical implications. In this regard, the stellar energy loss rates data have been used
to put constraints on the properties and interaction of light particles \cite{Dicus,Duane,Sthepen,Ellis}. In addition, one of the
most interesting possibilities to use stars as particle physics laboratories \cite{Raffelt1,Ruiz} is to study the backreaction of the
novel energy loss rates implied by the existence of new low-mass particles such as axions \cite{Payez,Fischer}, or by non-standard
neutrino properties such as magnetic moment and electric dipole moment \cite{Raffelt,Kerimov,Alexander,Blinnikov,Bugarin}. Our study
complements other studies on the dipole moments of the tau-neutrino at hadron and $e^+e^-$ colliders.

\vspace{8mm}

\begin{center}
{\bf Acknowledgments}
\end{center}

We acknowledge support from CONACyT, SNI and PROFOCIE (M\'exico).

\vspace{2cm}

%\newpage

\newpage

\begin{figure}[t]
\centerline{\scalebox{0.7}{\includegraphics{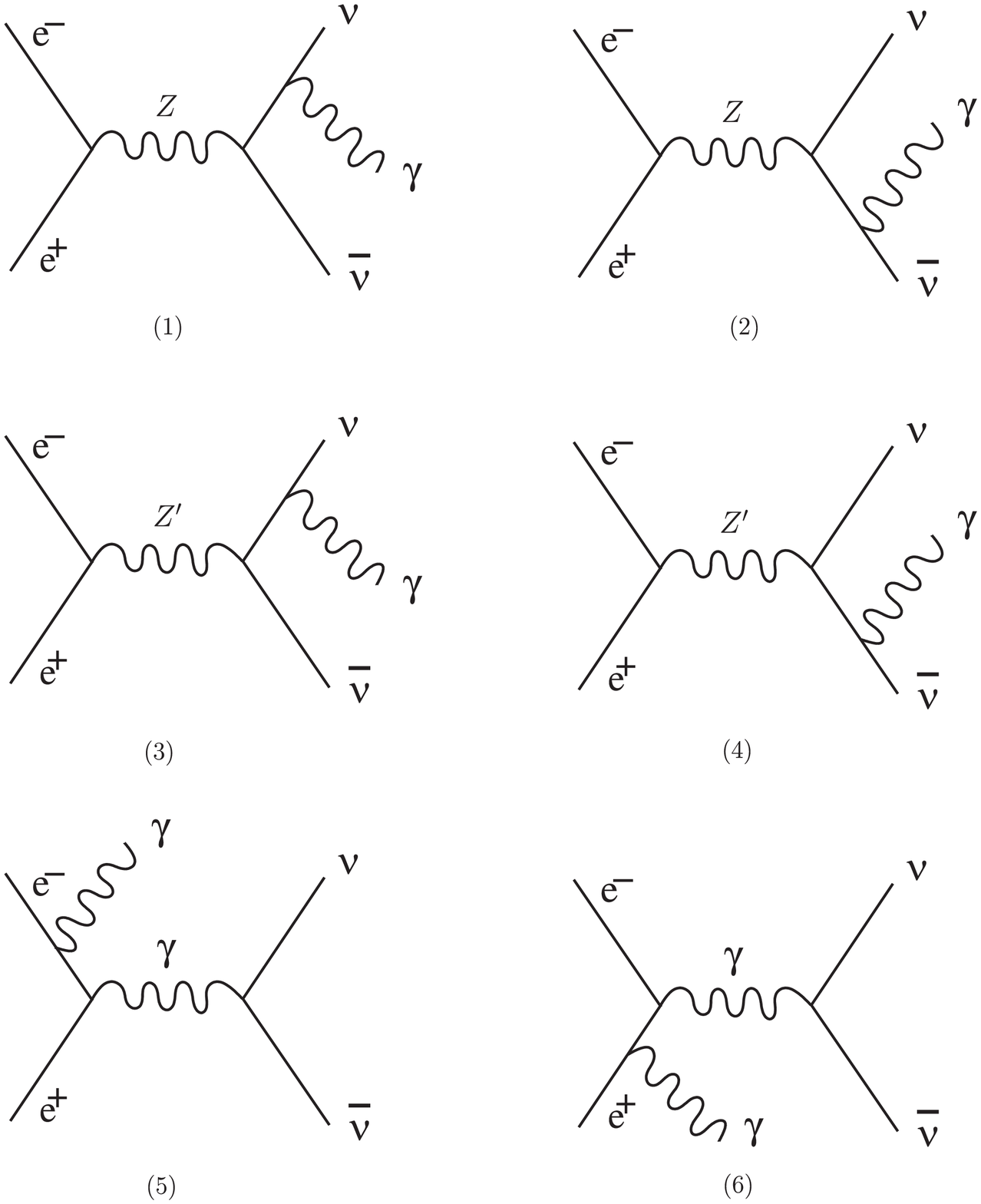}}}
\caption{ \label{fig:gamma} The Feynman diagrams contributing to
the process $e^{+}e^{-}\rightarrow \nu_\tau\bar\nu_\tau\gamma$ (1-4) when the
$Z(Z')$ vector bosons are produced on mass-shell and (5, 6) contributions from
anomalous neutrino electromagnetic couplings with initial-state radiation.}
\end{figure}

\begin{figure}[t]
\centerline{\scalebox{0.9}{\includegraphics{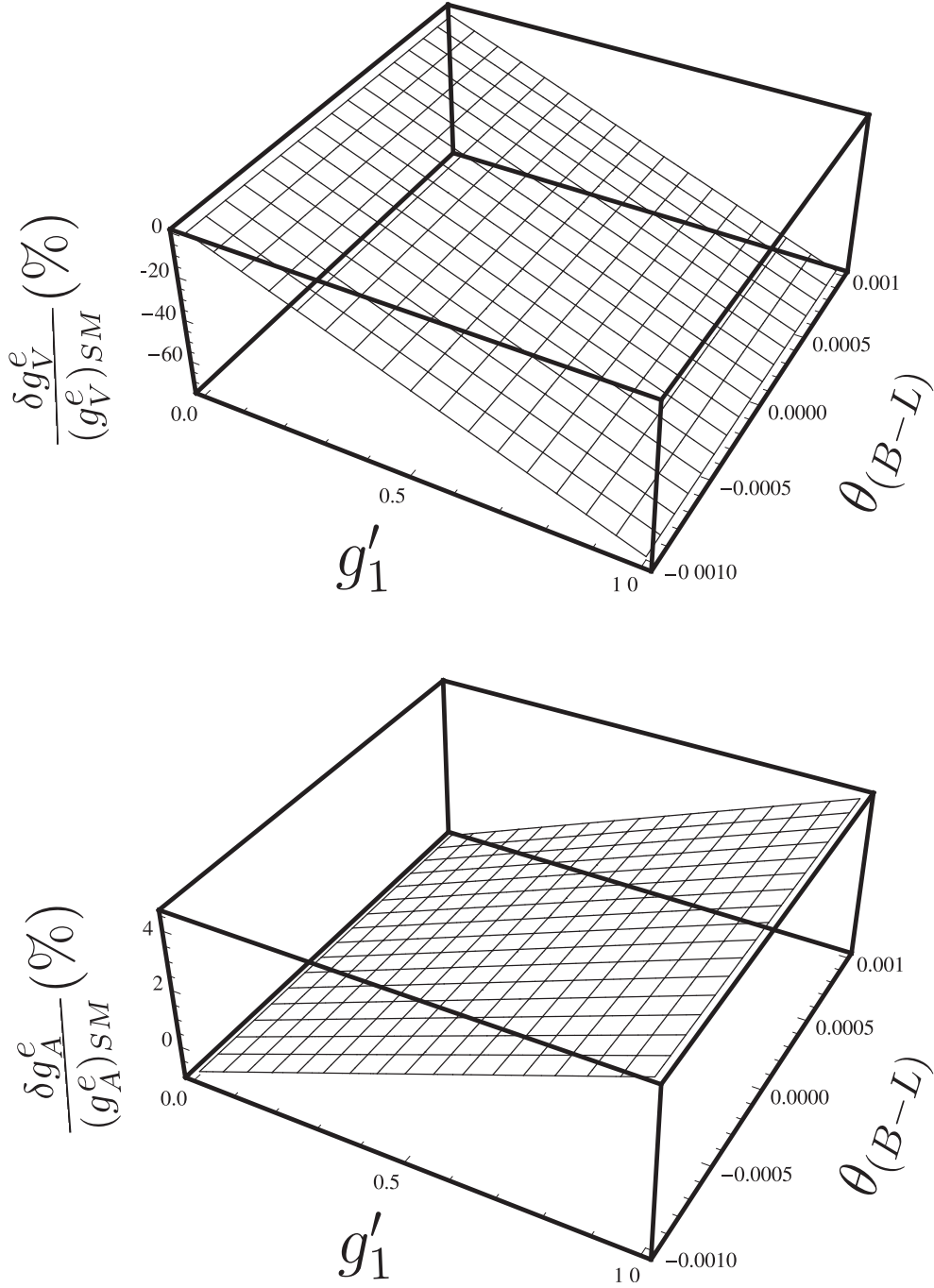}}}
\caption{ \label{fig:gamma} Top panel: The relative correction $\frac{\delta g^e_V}{(g^e_V)_{SM}}$
as a function of $(g'_1, \theta_{B-L})$. Bottom panel: The relative correction $\frac{\delta
g^e_A}{(g^e_A)_{SM}}$ as a function of $(g'_1, \theta_{B-L})$.}
\end{figure}

\begin{figure}[t]
\centerline{\scalebox{0.75}{\includegraphics{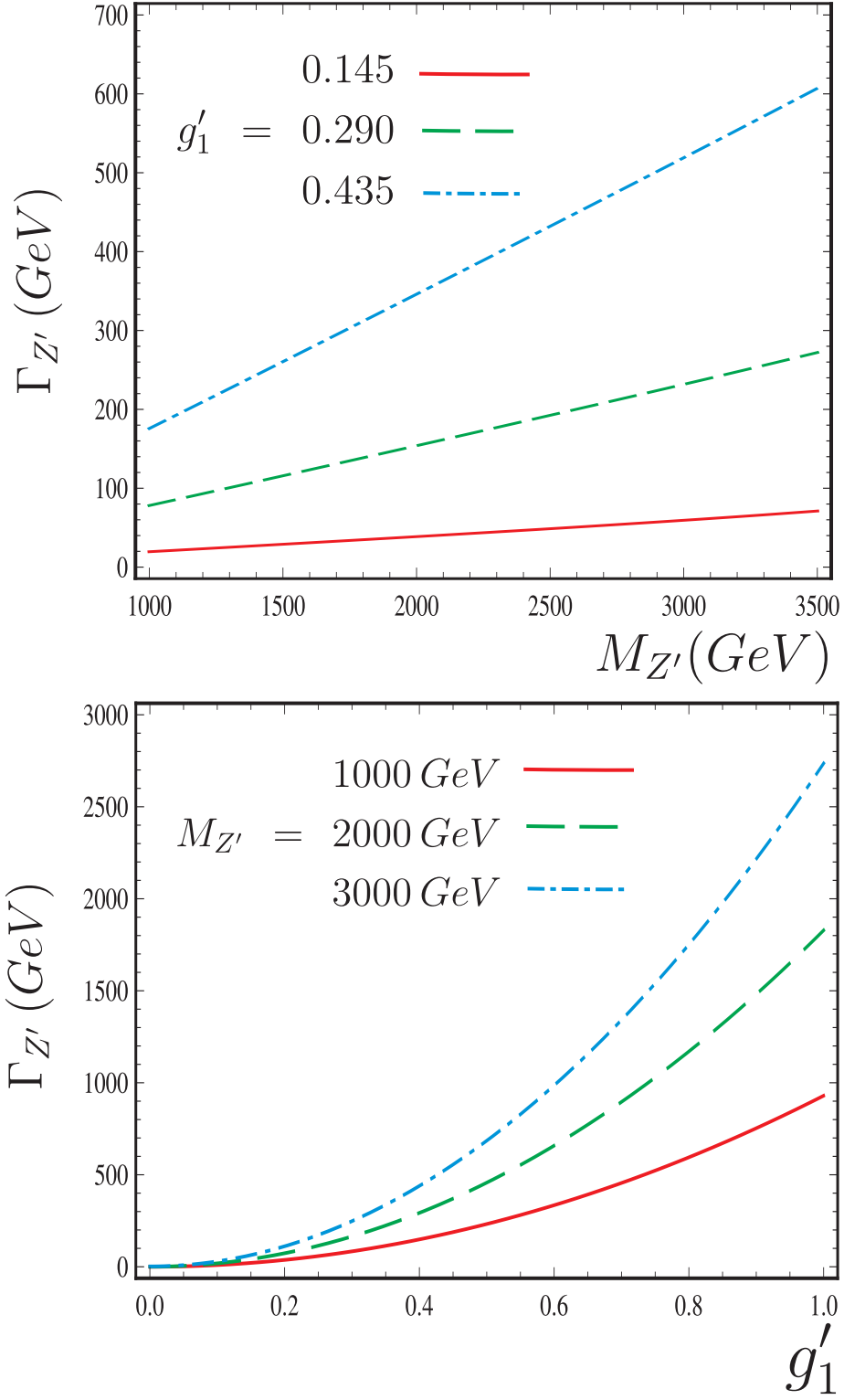}}}
\caption{ \label{fig:gamma} Top panel: $Z'$ width as a function of $M_{Z'}$
for fixed values of $g'_1$. Bottom panel: $Z'$ width as a function of $g'_1$
for fixed values of $M_{Z'}$.}
\end{figure}

\begin{figure}[t]
\centerline{\scalebox{0.67}{\includegraphics{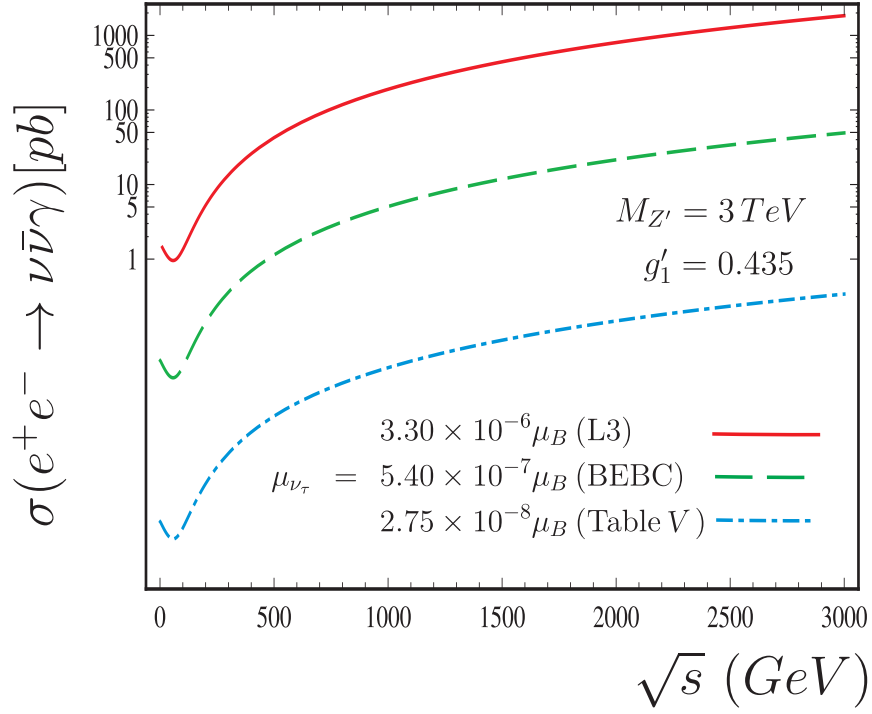}}}
\caption{ \label{fig:gamma} The curves show the shape for
$e^{+}e^{-}\rightarrow \nu_\tau \bar\nu_\tau \gamma$ as a function of
center-of-mass energy and different values of the $\mu_{\nu_\tau}$ magnetic moment.}
\end{figure}

\begin{figure}[t]
\centerline{\scalebox{0.67}{\includegraphics{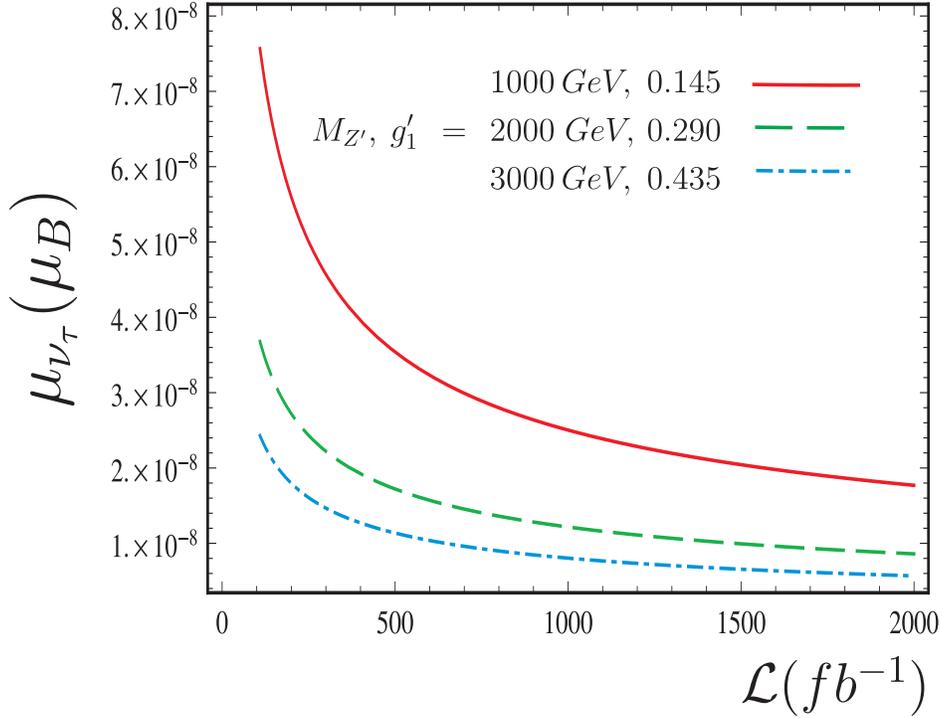}}}
\caption{ \label{fig:gamma} Dependence of the sensitivity limits at $95\%$ C. L. for the
anomalous magnetic moment for three different values of $M_{Z'}$ and $g'_1$ in the process
$e^{+}e^{-}\rightarrow \nu_\tau \bar\nu_\tau \gamma$.}
\end{figure}

\begin{figure}[t]
\centerline{\scalebox{0.75}{\includegraphics{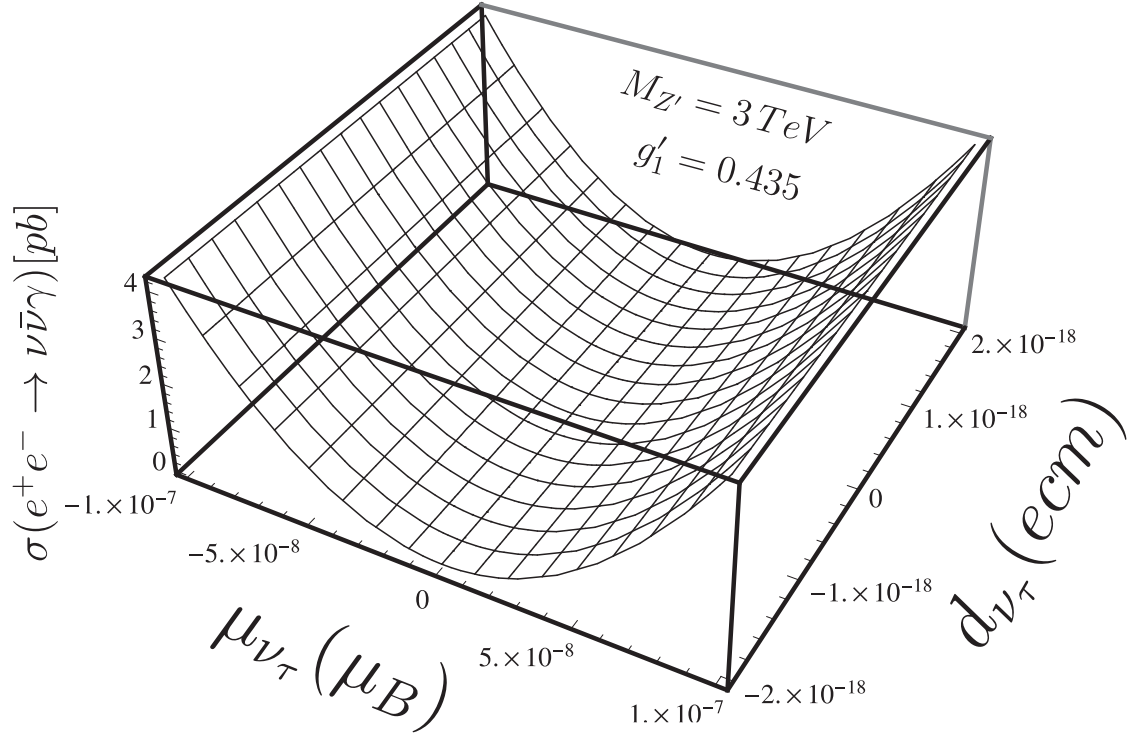}}}
\caption{ \label{fig:gamma} The surface show the shape for the cross section
of the process $e^{+}e^{-}\rightarrow \nu_\tau \bar\nu_\tau \gamma$ as a function
of the $\mu_{\nu_\tau}$ magnetic moment and the $d_{\nu_\tau}$  electric dipole moment.}
\end{figure}

\end{document}